# Visualizing improved spin coupling in molecular magnets

Judith Donner[1,2], Jan-Philipp Broschinski[3], Bastian Feldscher[3], Anja Stammler[3], Hartmut Bögge[3], Thorsten Glaser,[3*] and Daniel Wegner[1,2*]

[1] *Institute for Molecules and Materials, Radboud University, Heyendaalseweg 135, 6525 AJ Nijmegen, Netherlands.*

[2] *Physikalisches Institut and Center for Nanotechnology, Westfälische Wilhelms-Universität Münster, 48149 Münster, Germany.*

[3] *Lehrstuhl für Anorganische Chemie I, Fakultät für Chemie, Universität Bielefeld, Universitätsstr. 25, 33615 Bielefeld.*

[*]*email: D.Wegner@science.ru.nl, thorsten.glaser@uni-bielefeld.de*

January 13, 2016

A key to building functional devices on the basis of single molecule magnets in the framework of molecular electronics is the ability to deposit and study these molecules on a surface, because the structural, electronic and magnetic properties of molecules can significantly change upon adsorption. We have used the submolecular resolution of a scanning tunneling microscope to probe the local interactions within two different rationally designed single-molecule-magnet building blocks. A careful analysis of single-molecule spectroscopic maps reveals that the electronic properties are sensitively dependent on the molecular structure so that even small changes can drastically enhance or reduce the intramolecular spin coupling. Due to their planar geometry, these molecules are ideal model systems to study molecular magnetism of surface-supported complexes via scanning probe techniques.

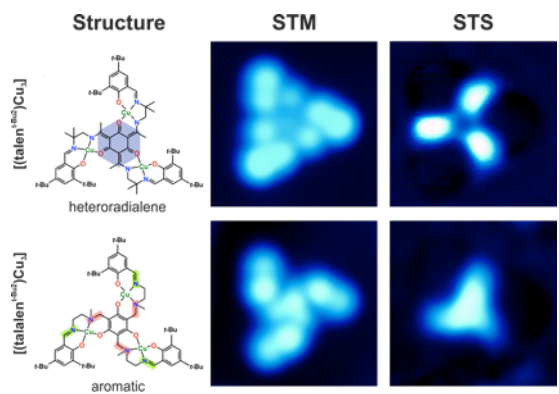

Single-molecule magnets (SMMs) are a promising class of nanomagnets due to chemical tunability of various properties and interactions. They exhibit a very slow relaxation of the magnetization which is caused by an energy barrier for magnetization reversal ($U^{eff}$) below a blocking temperature $T_B$[1]. This phenomenon has attracted increasing attention over the last two decades[2-9] and fueled speculations on potential applications[10-13]. However, all these applications are thought to require *i*) higher $T_B$'s and *ii*) a controlled deposition on surfaces with single-molecule addressability[14-17]. Besides the difficulties for a controlled deposition of SMMs on surfaces, studies on various molecular magnetic systems like the archetype SMM **Mn$_{12}$**[18-20] or on Co- and Fe-phthalocyanine molecules[21-24] showed that the surface can significantly alter the SMM properties via hybridization, charge transfer and ligand-field modifications. This can improve but also destroy SMM properties, as is the case for **Mn$_{12}$**. Understanding these influences calls for careful studies of adsorbed SMMs, for which scanning tunneling microscopy (STM) and spectroscopy (STS) are ideally suited.

Herein, we present our combined efforts to rationally increase the ferromagnetic interactions in our triplesalen-based complexes, to address the molecules on a surface individually by STM, and to study their local electronic properties via STS. Due to their planar molecular structure, every part of the molecules can be accessed, making them ideal candidates to study molecular magnetism via STM/STS.

Indeed, our local electronic structure analysis, as gathered by STM and STS, shows evidence that the rational improvement of the magnetic coupling is preserved for the adsorbed molecule. Thus, combining the rational improvement of our ligand system with the surface deposition and single-molecule addressability provides a further step towards fundamental understanding as well as future applications of SMMs.

Many studies on improving SMMs focus only on increasing $U^{eff}$ to slow down thermal relaxation over the top of the barrier. However, due to their quantum nature, SMMs can also tunnel through the energy barrier (quantum tunneling of the magnetization, QTM). We have focused our attention on minimizing QTM by a combination of a high spin ground state with a high molecular and crystal symmetry. In this respect, we have rationally developed a class of heptanuclear SMMs **[Mn$^{III}_6$M$^{III}$]$^{3+}$** (M = Cr$^{III}$, Mn$^{III}$) based on the triplesalen ligand H$_6$talen$^{t\text{-}Bu_2}$ [25-30]. The triplesalen ligand was designed to promote ferromagnetic interactions by the spin-polarization mechanism[31]. However, magnetic measurements of the building block [(talen$^{t\text{-}Bu_2}$)M$_3$] revealed that for M = Cu$^{II}$, the coupling between the Cu ions is indeed ferromagnetic but only weak[25,32] and for M = Mn$^{III}$ even weakly antiferromagnetic [27-29], which decreases $U^{eff}$ in our **[Mn$^{III}_6$M]$^{3+}$** SMMs.

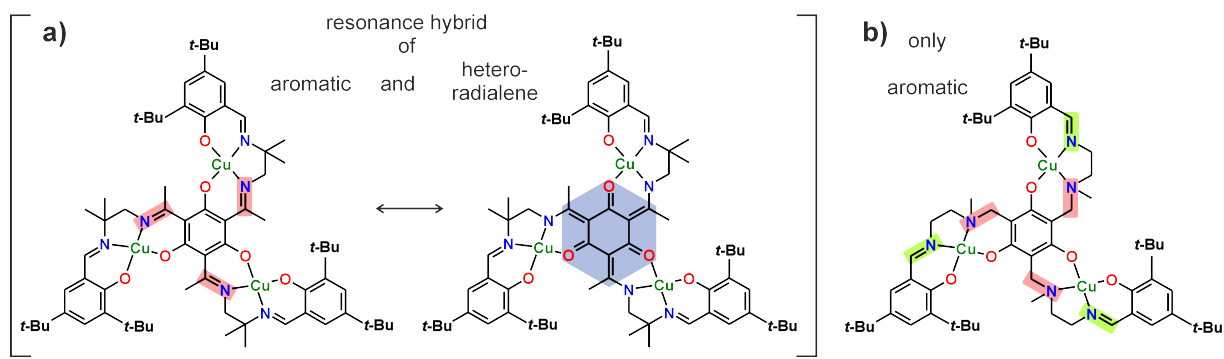

**Figure 1** a) Triplesalen complex [(talen$^{t\text{-Bu}_2}$)Cu$^{II}_3$]: the electronic structure is described by a resonance hybrid of an aromatic (left) and a non-aromatic heteroradialene (right) resonance structure (highlighted in blue). b) Triplesalalen complex [(talalen$^{t\text{-Bu}_2}$)Cu$^{II}_3$]: the electronic structure is described only by the aromatic resonance structure. The red accentuation highlights the change from an unsaturated C=N imine bond to a saturated C-N amine bond, which precludes heteroradialene formation. The green accentuation highlights the labile terminal imine bonds.

In order to understand why the triplesalen ligands do not efficiently promote ferromagnetic interactions by the spin-polarization mechanism, we analyzed spectroscopic and structural features of our compounds carefully and found that the triplesalen complexes are most likely not in the purely aromatic form which is needed for the spin-polarization mechanism but seem to be a resonance hybrid of the intended aromatic form (Figure 1a, left) and a non-aromatic heteroradialene form (right). This has no delocalized central ring that promotes the interaction between the three metal-subunits which in turn could prevent an effective spin-polarization[33]. As this heteroradialene can only be formed in the presence of unsaturated C=N imine groups (highlighted in red in Figure 1a), a rational improvement would be the replacement of the unsaturated C=N imine bonds by saturated C-N amine bonds (highlighted in red in Figure 1b). This design results in the new triplesalalen ligand H$_6$talalen$^{t\text{-Bu}_2}$ and its complex [(talalen$^{t\text{-Bu}_2}$)Cu$^{II}_3$] (Figure 1b), for which we have established new synthetic pathways (for further details concerning the design strategy and synthesis see Supporting Information Sections I-II). During the synthesis and crystallization, we also observed fragmentation, demonstrating that the remaining terminal imine groups of H$_6$talalen$^{t\text{-Bu}_2}$ are unstable towards hydrolysis even in the complexed form.

We have evaluated the heteroradialene character of this new ligand via NMR and optical spectroscopy as well as X-ray diffraction and found evidence that the triplesalen complex does not exhibit heteroradialen character (see Supporting Information Sections IV-V). Magnetic measurements indeed revealed a stronger coupling by about a factor of three for crystals of [(talalen$^{t\text{-Bu}_2}$)Cu$^{II}_3$] (compare Supporting Information Section VII).

However, as the structural, electronic and magnetic properties of molecules can significantly change upon adsorption it is essential to test whether the specific properties of these SMM building blocks are preserved on a surface. Therefore, we deposited the complexes [(talen$^{t\text{-}Bu2}$)Cu$_3$] and [(talalen$^{t\text{-}Bu2}$)Cu$_3$] onto a clean Au(111) surface for a direct structural and electronic characterization of the adsorbed molecule. To lift the central part of the molecule from the surface and thereby reduce the interaction with the substrate, we used bulky *tert*-butyl groups attached to the triplesalen ligand. Unfortunately, a standard thermal sublimation was not feasible due to the fragility of the large complexes. We therefore utilized the unconventional pulse injection technique[34,35] to deposit the molecules under ultrahigh vacuum conditions. After the preparation, the sample was immediately transferred into the cryogenic STM (operated at $T$ = 5 K) and topography images were taken to identify single molecules and to acquire spectroscopic maps of the molecular electronic structure. For details concerning the sample preparation and characterization compare the Supporting Information Sections IX-X.

Fig. 2 presents STM images of both molecules on Au(111). The topography of [(talen$^{t\text{-}Bu2}$)Cu$_3$] (a) shows threefold-symmetric adsorbates that are equally distributed over the surface. The adsorbates tend to lie in the more reactive regions of the Au(111) herringbone reconstruction, namely fcc regions and elbow sites. We do not observe any other adsorbates on the surface, which speaks for a clean sample preparation obtained by the pulse-injection technique. Fig. 2b shows a highly resolved topography image of a single [(talen$^{t\text{-}Bu2}$)Cu$_3$] molecule. We observe nine lobes with different intensities surrounding a depression in the center of the adsorbate. By comparing the adsorbates with the chemical structure (see Fig. 3a), they can be unambiguously identified as intact [(talen$^{t\text{-}Bu2}$)Cu$_3$] molecules. Apparently, the topography is dominated by the six *tert*-butyl and the three methyl groups, which are bulky and stick out farthest into the vacuum, thus dominating the tunneling process[36-38].

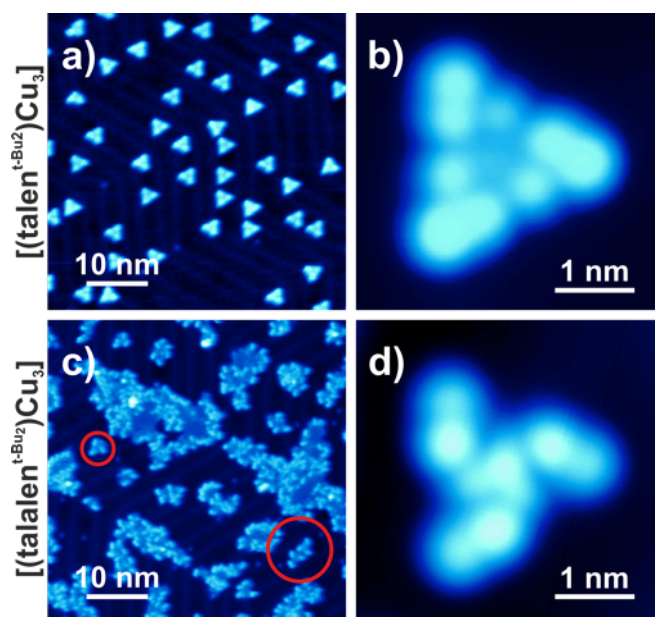

**Figure 2** a) [(talen$^{t\text{-Bu}_2}$)Cu$_3$] deposition via the pulse-injection technique leads to clean samples with intact isolated molecules. b) Sub-molecularly resolved STM images show adsorbates with a characteristic triangular shape which is in very good agreement with the molecular structure (See also Fig. 3a.). c) [(talalen$^{t\text{-Bu}_2}$)Cu$_3$] tends to dissociate, leaving islands of fragments and solvent molecules. d) However, also on this sample, we are able to find unbroken single molecules (cf. marked circles in c). (See also Fig. 3e.). Apparent heights depend on the applied sample-bias voltage and are between 0.27 and 0.41 nm for [(talen$^{t\text{-Bu}_2}$)Cu$_3$] and between 0.25 and 0.31 nm for [(talalen$^{t\text{-Bu}_2}$)Cu$_3$]. Image sizes: 49 x 49 nm$^2$ (a,c) and 3.7 x 3.7 nm$^2$ (b,d). All images were taken in constant current mode at a bias voltage of 1 V and a tunneling current of 50 pA.

In contrast, the overview image of [(talalen$^{t\text{-Bu}_2}$)Cu$_3$] molecules (Fig 2c) exhibits many different structures on the surface, mostly irregular and disordered clusters and islands of different sizes. This is consistent with the above described hydrolytic instability of the terminal imine groups even in the complexed form (for details of the sample topography, see the Supporting Informations Section XI). Nevertheless, we regularly found threefold symmetric adsorbates on the bare Au(111) surface in-between these islands. Close-up views of these adsorbates (as shown in Fig. 2d) reveal nine lobes dominating the topography, which is reminiscent to [(talen$^{t\text{-Bu}_2}$)Cu$_3$]. By overlaying the molecular structure of [(talalen$^{t\text{-Bu}_2}$)Cu$_3$] onto the topography (cf. Fig. 3e), we find a very good agreement and therefore conclude that the threefold symmetric adsorbates are intact [(talalen$^{t\text{-Bu}_2}$)Cu$_3$] molecules.

Despite a certain degree of reminiscence, a direct comparison of the topography features between [(talen$^{t\text{-Bu}_2}$)Cu$_3$] and [(talalen$^{t\text{-Bu}_2}$)Cu$_3$] already shows significant distinctions, regardless of the fact that the molecular structures have only seemingly subtle differences. For a more thorough analysis of the molecular electronic properties, we performed STS-based energy-dependent spectroscopic mapping. In brief, these spectroscopic (or *dI/dV*) maps reflect the spatial distribution of the local density of states at the selected voltage, which essentially allows for imaging molecular orbitals (i.e. |Ψ|$^2$)[39-41]. Negative (positive) voltages correspond to (un)occupied states, and *V* = 0 defines the Fermi energy. We found clear resonances in both the occupied and unoccupied energy regions, as summarized in the *dI/dV* maps shown in Fig. 3 (for a complete set of *dI/dV* maps, compare the Supporting Informations Section XII).

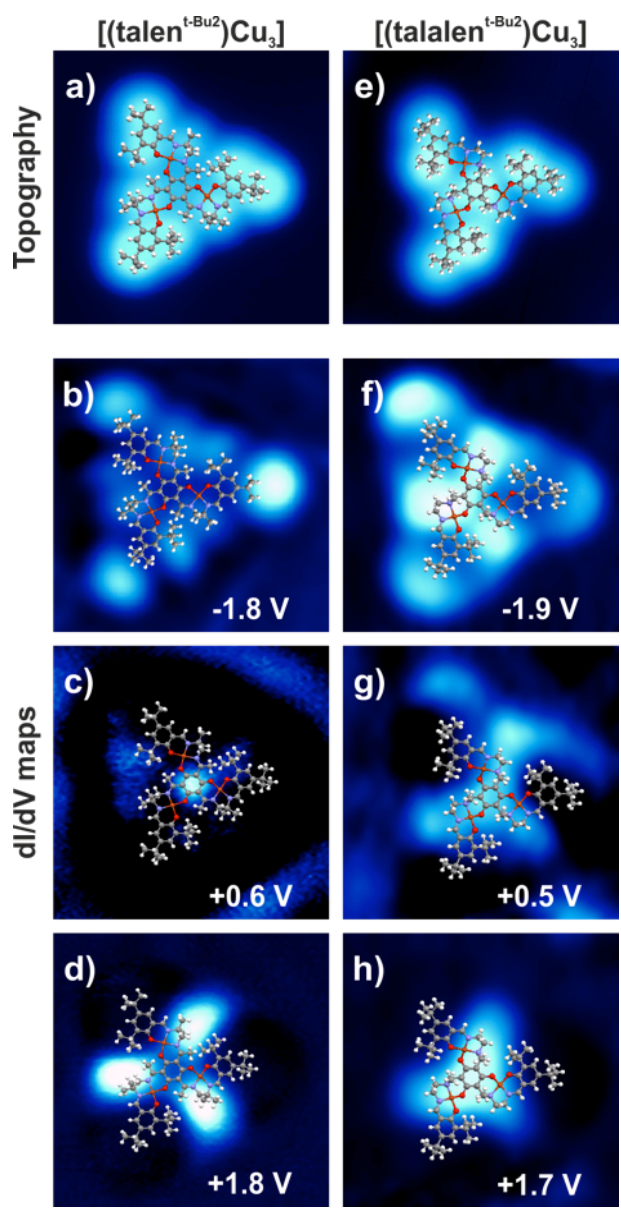

**Figure 3** a) STM image of a single [(talen$^{t\text{-Bu}_2}$)Cu$_3$] molecule with its molecular structure overlaid. b)-d) corresponding *dI/dV* maps. e) STM image of a single [(talalen$^{t\text{-Bu}_2}$)Cu$_3$] molecule, and f)-h), corresponding *dI/dV* maps. We find a clear difference in the orbital distribution, especially when focusing on the central carbon ring that connects the three salen subunits. Image sizes are 3.7 x 3.7 nm$^2$. All dI/dV maps were taken at a set point current of 50 pA using lock-in detection (modulation voltage 20 mV rms, frequency ≈ 750 Hz).

We first focus on maps of [(talen$^{t\text{-Bu}_2}$)Cu$_3$]. In the occupied region, we observed resonances located at each of the *tert*-butyl and methyl groups (Fig. 3b). They are visible as lobes that are separated from each other by nodal planes. In the center of the molecule (i.e., at the central carbon ring) we found no resonance over the entire accessible occupied energy range ($E - E_F$ = -2.4…0 eV). However, in the unoccupied region in an energy window 0.6…1.2 eV, we found a pronounced intensity localized exclusively on the central carbon ring, while the rest of the molecule remains dark (c). Only above 1.2 eV, when the central feature starts to disappear, additional intensity arises along the edges of the triangle-shaped molecule, thus forming a windmill-like pattern (d). This latter feature remains visible even above 2 eV, where it becomes a very strong resonance. Hence, our spectroscopic maps show that the central carbon ring seems to be mostly electronically decoupled from the rest of the molecule. Indeed, there is only a very small energy region ($E - E_F = 1.2…1.4$ eV) where we observed simultaneous intensity on central and other parts of the molecule.

In comparison, $dI/dV$ maps of [(talalen$^{t\text{-Bu2}}$)Cu$_3$] reveal a different behavior. In the occupied region, there is clearly intensity at the central carbon ring and the edge region of the triangular molecule (Fig. 3f). The differences are even better visible in the unoccupied region: for $E - E_F > 0.3$ eV, we observed the windmill-like pattern but also here we found a simultaneous high intensity at the central ring (g). In a similar fashion, this feature becomes a very strong resonance at around 2 eV, but again the high density of states remains on the molecular center, too (h). Overall, a comparison between [(talen$^{t\text{-Bu2}}$)Cu$_3$] and [(talalen$^{t\text{-Bu2}}$)Cu$_3$] reveals that the changes in the molecular structure do not affect the outer *tert*-butyl phenol groups, but the electronic structure changes drastically on the central part of the molecule. Most importantly, for [(talalen$^{t\text{-Bu2}}$)Cu$_3$] all observed molecular orbitals involve local density of states at the central carbon ring, whereas in the case of [(talen$^{t\text{-Bu2}}$)Cu$_3$] the molecular center behaves quite independently from the rest of the molecule.

All this suggests that in [(talen$^{t\text{-Bu2}}$)Cu$_3$] the orbitals involving the central carbon ring are not hybridized with the ones of the rest of the molecule. Instead, the molecular center appears as a discontinuity in the orbital distribution of the three salen subunits so that there is no real electronic connection between them. This can effectively weaken the coupling between the three metal centers and thus the spin coupling[42]. In contrast, this behavior is not present for any of the [(talalen$^{t\text{-Bu2}}$)Cu$_3$] orbitals. All three salen subunits are always connected through the central ring, i.e., all molecular orbitals are also more delocalized over the entire molecule. As a strong orbital delocalization is a prerequisite for an efficient high spin-polarization over the central ring[43-45], this is indicative of an enhanced intramolecular coupling of the metal centers. To summarize the comparison, our STM measurements provide clear indications that indeed the central carbon ring in [(talen$^{t\text{-Bu2}}$)Cu$_3$] is best described as a heteroradialene-like structure with exocyclic double, while [(talalen$^{t\text{-Bu2}}$)Cu$_3$] exhibits a benzene-like delocalized aromatic π-system that leads to the strengthening of the spin coupling via highly delocalized molecular orbitals. Hence, we find that the surface does not significantly affect the intrinsic electronic structure of the [(talen$^{t\text{-Bu2}}$)Cu$_3$] and [(talalen$^{t\text{-Bu2}}$)Cu$_3$] complexes, which shows that also the specific intramolecular spin coupling is preserved upon adsorption.

In conclusion, our combined chemical synthesis and STM/STS study shows that a careful analysis of the chemical structure of SMM building blocks and its corresponding orbitals allows to identify, what determines the coupling of spins in a molecular magnet. As our STM measurements reveal that the spin coupling in [(talen$^{t\text{-Bu2}}$)Cu$_3$] and [(talalen$^{t\text{-Bu2}}$)Cu$_3$] complexes is preserved for the adsorbed molecules we suggest that also single molecule magnets from that family will retain their magnetic properties, which would be an important step for a future device-integration. Moreover, while for many other SMMs an STM-based analysis is challenging due to difficulties in identifying the molecular adsorption geometry of the bulky molecules an advantage of the molecules presented here is their planar molecular structure which makes them ideal model systems to study molecular magnetism via scanning probe-based experiments. We are convinced that our findings also pave the way for an improvement of the rational design strategy where synthetic approaches to enhance the spin coupling are complemented by the spatial resolution of scanning probe techniques to reveal effects of molecule-surface interactions.

**Acknowledgments**

We thank the Deutsche Forschungsgemeinschaft (DFG) for financial support through the Research Unit FOR 945 (project P1) as well as project WE 4104/2-1. J.D. acknowledges support by the Studienstiftung des Deutschen Volkes.

**Additional information**

Supporting information is available. Correspondence should be addressed to T.G. and D.W.

**Competing financial interests**

The authors declare no competing financial interests.